Understanding the role of surface plasmon polaritons in two-dimensional achiral nanohole arrays for polarization conversion

Z.L. Cao[1], L.Y. Yiu[1], Z.Q. Zhang[2], C.T. Chan[2], and H.C. Ong[1, a]

1) Department of Physics, The Chinese University of Hong Kong, Shatin, Hong Kong, People's Republic of China

2) Department of Physics, The Hong Kong University of Science and Technology, Clear Water Bay, Hong Kong, People's Republic of China

We have studied the dependence of the rotation angle and ellipticity on the sample orientation and incident polarization from metallic nanohole arrays. The arrays have four-fold symmetry and thus do not possess any intrinsic chirality. We elucidate the role of surface plasmon polaritons (SPPs) in determining the extrinsic chirality and we verify the results by using finite-difference time-domain simulation. Our results have indicated the outgoing reflection arises from the interference between the nonresonant background, which preserves the input polarization, and the SPP radiation damping, which is linearly polarized but carries a different polarization defined by the vectorial field of SPPs. More importantly, the interference manifests various polarization states ranging from linear to elliptical across the SPP resonance. We analytically formulate the outgoing waves based on temporal coupled mode theory (CMT) and the results agree well with the experiment and simulation. From CMT, we find the polarization conversion depends on the interplay between the absorption and radiative decay rates of SPPs and the sample orientation.

a) hcong@phy.cuhk.edu.hk

## I. INTRODUCTION

Polarization is one of the most fundamental parameters of electromagnetic waves and it defines many intriguing optical phenomena [1]. Therefore, how one can manipulate the polarization state has been a major concern not only from a scientific point of view but also from a practical consideration. Conventional methods rely primarily on using birefringent materials that have anisotropic refractive index [2]. Half- and quarter-wave plates are two prominent examples that either rotates a linearly polarized light or converts it into a circular polarization [2]. With the emergence of nanophotonics, materials can now be designed at the length scale of nanometers to engineer different wave properties including polarization. Photonic crystals [3,4], plasmonic systems [5-9], metamaterials [10-17], and metasurfaces [18-23] have been reported to control the polarization state at different extents. In the early works of plasmonic systems, birefringent-like environment are created by using elliptical nanoholes or nanoparticles in periodic lattices that break the space invariance or mirror symmetry when the major axis of the basis is tilted away from the incident polarization [9]. Since then, this symmetry breaking technique has been widely applied to design various shapes of the basis in plasmonic systems and metamaterials for polarization conversion. Gammadion [10,24-25], spiral [26,27], helix [28,29], cross [30,31], L-, G-, and S-shape [32-34], etc, have been extensively studied to exhibit various degrees of optical activity. These entities induce strong chiral near fields that evolve into different polarization states. Other than the intrinsic chirality, extrinsic chiral effects are drawing attention as well. For example, nonlocal effect has been reported to control polarization [35]. Polarization conversion can occur in achiral metallic arrays enabled by spatial dispersion [36]. The nonlocality induces anisotropic optical responses along and out of the incident plane, leading to birefringence. In addition, achiral metamaterials have shown strong optically activity if the incident light and the sample orientation form a chiral triad that breaks symmetry [37,38]. Surprisingly, propagating surface plasmon polaritons (SPPs) have recently renewed the interest in extrinsic chirality. It is observed that under certain excitation condition, SPPs from achiral systems can produce much stronger circular dichroism than the gammadion metamaterials [39-41]. Therefore, a complete understanding of the effects of SPPs on polarization conversion is necessary for getting better control. However, while SPPs have been reported to yield polarization conversion for more than twenty years, the underlying physics is not yet fully understood [7,42,43].

In this work, we have studied the rotation angle ($\psi$) and ellipticity ($\chi$) from two-dimensional (2D) square lattice circular nanohole arrays by using angle- and polarization-resolved reflectivity spectroscopy. Our results demonstrate SPPs play a significant role in controlling the polarization state of the outgoing wave. In particular, both $\psi$ and $\chi$ indicate the polarization state exhibits a very complicated behavior, spanning from almost circular to linear polarization when crossing the SPP resonance. The experimental results are verified by finite-difference time-domain (FDTD) simulations. Furthermore, we find the polarization is determined by the interference between the nonresonant reflection that contains the same polarization as the incidence and the resonant SPP radiation damping in which the polarization is defined by the vectorial near field of SPPs. To support this, we have analytically formulated the outgoing polarization based on temporal coupled mode theory (CMT) [44] and the results agree well with the experiment and simulation. The theory stresses the importance of the interplay between the absorption and radiative decay rates of SPPs and the sample orientation in polarization conversion.

## II. EXPERIMENTAL METHODS

2D square lattice gold (Au) nanohole arrays are fabricated by interference lithography as described earlier [45]. The scanning electron microscopy (SEM) image of one sample is illustrated in the inset of Fig. 1(a) as an example, showing it has period P = 800 nm, hole depth H and radius R = 100 and 116 nm. The structure possesses a four-fold symmetry and it is thus achiral. Since the Au film is optically thick, the sample has no transmission. After sample preparation, it is placed on a computer-controlled goniometer for angle- and polarization-resolved reflectivity spectroscopy [46]. The setup is shown in Fig. 1(a). White light from a quartz lamp is collimated and weakly focused onto the sample at a well-defined incident angle θ. The sample can be rotated with respect to the surface normal for different azimuthal angles φ defined as the angle between the incident plane and the Γ-X direction of the lattice. An incident polarizer is located between the light source and the sample whereas a quarter-wave plate and an analyzer can be placed after the sample for polarimetric measurements. The specular reflections are collected by a CCD detector attached to a spectrometer. By contour measuring the reflectivity at different θ and φ, one can map out the dispersion relations of the arrays for mode identification [45,46]. At the same time, the polarization state of the outgoing reflection can be accessed by measuring both ψ and χ [47]. In general, ψ and χ are given as $tan\,2\psi = S_2/S_1$ and $sin\,2\chi = S_3/S_0$, where $S_{0-3}$ are the four Stokes parameters. The parameters are related to the reflection intensities $I$ as

$S_0 = I(0°,0°) + I(90°,0°)$, $S_1 = I(0°,0°) - I(90°,0°)$,

$S_2 = 2I(45°,0°) - I(0°,0°) - I(90°,0°)$, and $S_3 = 2I(45°,90°) - I(0°,0°) - I(90°,0°)$, where the parenthesis (γ,β) defines the orientation of the analyzer and the phase retardation introduced by the quarter wave plate [47]. The transmission axis of analyzer can be set at γ = 0°, 45°, and 90° with respect to the incident plane by either removing the quarter wave plate (i.e. β = 0°) or inserting the wave plate with the fast axis parallel to γ = 0° (i.e. β = 90°) (see Fig. 2(a)) [47]. Therefore, the reflections at four detection configurations, $I(0°,0°)$, $I(90°,0°)$, $I(45°,0°)$, and $I(45°,90°)$, allow one to determine all four Stokes parameters as well as ψ and χ.

## III. RESULTS

### a. Angle-dependent reflectivity, rotation, and ellipticity mappings

We first show the φ-dependent p-polarized reflectivity mapping of the array in Fig. 1(b) taken at θ = 10°. From the mapping, two dispersive low reflection bands are seen and they are identified as two Bloch-like SPPs by using the phase-matching equation [45,46]:

$$\frac{2\pi}{\lambda_{SPP}}\sqrt{\frac{\varepsilon_{Au}}{\varepsilon_{Au}+1}} = \sqrt{\left(\frac{2\pi}{\lambda_{SPP}}\sin\theta\cos\varphi + \frac{2n\pi}{P}\right)^2 + \left(\frac{2\pi}{\lambda_{SPP}}\sin\theta\sin\varphi + \frac{2m\pi}{P}\right)^2}, \quad (1)$$

where $\varepsilon_{Au}$ is the dielectric constant of Au extracted from Ref [48], $\lambda_{SPP}$ is the SPP resonant wavelength, and (n,m) are the integers defining the order of SPPs. As indicated by the dash lines, Eq. (1) shows two (-1,0) and (0,-1) SPPs are excited. At φ = 45° where the SPPs cross, we see a small plasmonic band gap emerges together with the formation of a pair of hybridized dark and bright modes that feature with different radiation damping rates [49,50]. The dark mode located at longer wavelength is nonradiative and thus is barely seen while the bright mode is at shorter wavelength, displaying a strong reflection dip [49,50]. For the polarimetric measurements, Fig. 2(b) & (c) show the corresponding ψ and χ contour mappings. By comparing three mappings, we clearly see they are closely related. One can also see the non-resonant reflection background, in which the array acts as a flat mirror and thus has high reflectivity, does not induce any noticeable ψ and χ, evidently showing both ψ and χ are mediated by SPPs. When tracking along the (-1,0) SPP mode in the ψ mapping, for example, we see ψ decreases from zero to negative when φ increases, and then flips to positive at $\lambda_{SPP}$ ~ 940 nm (i.e φ = 20°). The signs are reversed for the (0,-1) SPPs. On the other hand, at any φ in the χ mapping, the χ of (-1,0) mode transits from positive to negative when scanning from short to long wavelength but becomes zero at $\lambda_{SPP}$. This trend is again reversed for the (0,-1) mode. At the cross point, both ψ and χ are almost zero.

To examine our results more carefully, we extract $\psi$ and $\chi$ as a function of $\varphi$ along the (-1,0) and (0,-1) modes in Fig. 2(d) – (f). For two modes, both $\psi$ and $\chi$ exhibit inversion symmetry in magnitude and sign, as expected from four-fold symmetry. For $\psi$ in Fig. 2(d), at the gap where $\varphi = 45°$, $\psi$ becomes zero. In addition, $\psi$ varies dramatically near the gap region, featuring an anomalous "oscillation" superimposed on the broad $\psi$ background. For $\chi$, we extract the largest positive and negative $\chi$ around $\lambda_{SPP}$ as well as the $\chi$ exactly at $\lambda_{SPP}$ for two modes and plot them in Fig. 2(e) & (f). In fact, $\chi$ is zero along $\lambda_{SPP}$. For the positive and negative $\chi$, similar "oscillation" features overlying on the broad backgrounds are seen at the gap region. By summarizing the behaviors of $\psi$ and $\chi$, one physically can imagine at $\lambda_{SPP}$ the outgoing wave is linearly polarized but the polarization is rotated away from the incident plane defined by $\psi$. However, when the wavelength is slightly off the $\lambda_{SPP}$, the reflection becomes right or left elliptically polarized depending on the mode order and wavelength. More importantly, an additional but unknown effect is involved, giving rise to the anomalies in both $\psi$ and $\chi$ around the gap region.

**b. Finite-difference time-domain simulation**

To verify our experimental results, we have conducted FDTD simulations. The unit cell is shown in the inset of Fig. 1(a) and it has P = 800 nm, H = 100 nm and R = 116 nm. A small modulation with height = 35 nm is added by referencing to the SEM. Bloch boundary condition is used on four sides and perfectly matched layer is used on the top and at the bottom [51]. At $\theta = 10°$, we calculate the p-polarized reflectivity, $\psi$, and $\chi$ mappings in Fig. 1(c) and 3(a) & (b). We find the calculation results agree well with the experiment. They all exhibit similar dependences of the magnitude and sign of $\psi$ and $\chi$ on the SPP modes. The theoretical $\psi$ for the (-1,0) and (0,-1) modes and the $\chi$ for the (0,-1) mode are plotted in Fig. 3(c) & (d) as a function of $\varphi$. $\chi$ is zero along the SPP modes, indicating linear polarization.

Both $\psi$ and $\chi$ are zero at the gap. At the gap region, similar anomalies appear although they look smaller and sharper. The only imperfection between FDTD and experiment is the flipping of $\psi$ is more extreme and the polarization state is completely converted from p to s at $\lambda_{SPP}$ = 950 nm (i.e. $\varphi \sim 20^o$) in simulation. In addition, when off the resonance, the reflection is almost circularly polarized in which $\chi$ is close to $\pm 45^o$.

## IV. RADIATION OF SPPs

### a. Dependence of SPP excitation on incident polarization angle

To elucidate the importance of SPPs in determining $\psi$ and $\chi$ and the occurrence of the anomalies at the gap region, one must first understand how SPPs are excited and then decay radiatively in periodic arrays. In particular, the polarization state of the SPP radiation damping is expected to play a key role in controlling the outgoing polarization. We have performed two types of experiments. The first measures the reflectivity as a function of incident polarization angle $\alpha$, defined with respect to the incident plane, at $\theta$ and $\varphi$ specifically for exciting a particular (-1,0) SPPs. $\alpha = 0^o$ and $90^o$ define the p- and s-incidences. No analyzer and quarter wave plate are used. One example is plotted in Fig. 4(a) for $\theta$ and $\varphi$ = $10^o$ and $10^o$, corresponding to the excitation of (-1,0) SPPs at 950 nm. It exhibits a sinusoidal-like behavior and the reflectivity minimum is located at $\alpha_{min}$ = $168^o$. Keeping $\theta = 10^o$ while changing $\varphi$, we see similar sinusoidal curves for other (-1,0) $\lambda_{SPP}$ but $\alpha_{min}$ is being shifted [Fig. 4(b)]. We then plot $\alpha_{min}$ as a function of $\lambda_{SPP}$ in Fig. 4(c), showing $\alpha_{min}$ increases gradually with $\lambda_{SPP}$ but diverges at ~ 910 nm where the gap is located (i.e. $\varphi$ = $45^o$) to $180^o$ and $90^o$ for the bright and dark modes. In fact, $\alpha_{min}$ can be interpreted as the best polarization angle for exciting SPPs, in which much of the energy is channeled to SPPs for yielding low reflectivity. Therefore, $\alpha_{min}$ implies the overlapping of the incident and the SPP electric fields is maximal so that the coupling between them is optimal [8]. In other words, as shown in the inset of Fig. 4(c), considering the incident polarization unit vector as $\hat{e}$ and the plasmonic field as $\vec{E}_{SPP}$, $\alpha_{min}$ occurs when $\hat{e} \cdot (\vec{E}_{SPP} \times \hat{z}) = 0$, where $\hat{z}$ is the unit vector normal to the surface, so that two fields lie on the same plane. In addition, for nondegenerate propagating SPPs where the longitudinal component of $\vec{E}_{SPP}$ is always parallel with the propagation direction $\hat{k}_{SPP}$, the above condition can be rewritten as $\hat{e} \cdot (\hat{k}_{SPP} \times \hat{z}) = 0$. Given $\hat{e} = \cos\alpha \cos\theta \hat{x} + \sin\alpha \hat{y} - \cos\alpha \sin\theta \hat{z}$ and $\hat{k}_{SPP} = \cos\rho \hat{x} + \sin\rho \hat{y}$, where ρ is the propagation angle defined with respect to the incident plane, the vector product yields:

$$\tan\alpha_{min} = \cos\theta \tan\rho. \quad (2)$$

In general, ρ is determined by rearranging the phase matching equation in Eq. (1) as:

$\rho = \tan^{-1}\left(\dfrac{\sin\theta \sin\varphi + n\lambda_{SPP}/P}{\sin\theta \cos\varphi + m\lambda_{SPP}/P}\right) + \varphi$. For (-1,0) SPPs, we calculate $\alpha_{min}$ for different $\lambda_{SPP}$ and plot it in Fig. 4(c) for comparison. We find it agrees with experiment except at the cross region. The deviation is due to the fact that at the cross point where two degenerate SPPs couple, they interfere and form two standing waves as $\vec{E}_{SPP}^1 + \vec{E}_{SPP}^2$ and $\vec{E}_{SPP}^1 - \vec{E}_{SPP}^2$ for the bright and dark modes [52,53]. The resulting electric field vectors thus point along and normal to the incident plane for two modes, leading to the product $(\vec{E}_{SPP}^1 + \vec{E}_{SPP}^2) \times \hat{z}$ and $(\vec{E}_{SPP}^1 - \vec{E}_{SPP}^2) \times \hat{z}$ that are perpendicular and parallel to the incidence. As a result, $\alpha_{min}$ is determined to be 180° and 90° for the bright and dark modes, in consistent with our results. Fig. 4(d) shows $\alpha_{min}$ as a function of $\lambda_{SPP}$ taken at different θ together with the analytical models for nondegenerate (-1,0) SPPs (i.e. exclude the cross regions). Except at θ = 15° where discrepancy is seen at short wavelengths, the good agreement between them verifies the condition for SPP excitation.

We also perform FDTD simulations to further confirm Eq. (2). First, we mimic our experiment to determine $\alpha_{min}$ by simulating the reflectivity as a function of $\alpha$ at $\theta = 10^\circ$ for different $\lambda_{SPP}$ and the results are plotted in Fig. 5(a). Second, we determine the propagation direction angle $\rho$ of the corresponding SPPs under the same $\theta$ by calculating the Poynting vector. The Poynting vector maps taken at two $\varphi = 30^\circ$ and $45^\circ$ for $\lambda_{SPP} = 936$ and 907.5 nm are shown in Fig. 5(b) & (c) for illustration. In the unit cell, the Poynting vector is determined by integrating the vectors at four boundaries. With both $\rho$ and $\theta$ ready, $\alpha_{min}$ is obtained from Eq. (2) and plotted in Fig. 5(a) for comparison. Despite some minor discrepancy, two independent methods give almost the same trend, validating Eq. (2). Therefore, considering the reciprocity theorem [54], we speculate that the polarization of the outgoing SPP radiation, defined as $\phi_{SPP}$ with respect to the incident plane, should follow $\alpha_{min}$ for any given sample orientation.

**b. Polarization angle of SPP radiation damping**

To prove the speculation, we conduct the second experiment. This time, we place the analyzer in the detection path and orient it so that the polarizer and analyzer are always perpendicular to each other. Therefore, the measured reflectivity contains no contribution from the nonresonant reflection but only the component of SPP radiation damping projected onto the transmission axis of the analyzer. Since the $\phi_{SPP}$ of the SPP radiation is always equal to $\alpha_{min}$, which remains unchanged provided the sample orientation is fixed, the orthogonal polarizer and analyzer pair only affects how much power is channeled to SPPs but not $\phi_{SPP}$. As an example, Fig. 6(a) shows the orthogonal reflection measured at $\theta$ and $\varphi = 10^\circ$ and $10^\circ$ (i.e. (-1,0) $\lambda_{SPP} = 950$ nm) as a function of $\alpha$, showing a sinusoidal behavior. Several more are taken at other $\varphi$ in Fig. 6(b), exhibiting similar sinusoidal but displaced curves. To find $\phi_{SPP}$, we refer to Fig. 6(c) for the outgoing wave, which shows the polarization of the SPP radiation together with the transmission axes for the polarizer and analyzer and the incident plane. Given the SPP radiation with intensity $I_{SPP}$ is linearly polarized at $\phi_{SPP}$, the signal after the analyzer is $I(\alpha) = I_{SPP}(\alpha)\cos^2(\phi_{SPP} - \gamma) = I_{SPP}(\alpha)\sin^2(\alpha + \phi_{SPP})$, where $\alpha + \gamma$ is always equal to $\pi/2$ for the orthogonal pair. Knowing from Fig. 4(b) that $I_{SPP}(\alpha)$ should follow a general sinusoidal function $A + B\sin(a\alpha + b)$ with all the capital and small A and B are constants, we fit Fig. 6(b) to determine $\phi_{SPP}$. The results of $\phi_{SPP}$ for different $\varphi$ are plotted in Fig. 6(d). The data from Fig. 4(c) is also superimposed on it, showing an almost perfect match to conclude $\alpha_{min} = \phi_{SPP}$.

**V. Coupled mode theory for the reflection interference**

Accordingly, the outgoing specular reflection is expected to carry two polarization components and they are the non-resonant reflection, which is solely determined by the incident polarization, and the SPP radiation damping, which is linearly polarized with the rotation determined primarily by the plasmonic field. This knowledge can then be transformed into analytical reflection coefficients by using temporal CMT [44,46,49,53]. Under p-polarized excitation at fixed $\theta$ and $\varphi$, the transient of SPP mode amplitude $a$ can be written as:

$$da/dt = i\omega_{SPP}a - \Gamma_{tot}a/2 + \sqrt{\Gamma_{rad}}\, e^{i\delta} s_+ \cos\alpha_{min}, \qquad (3)$$

where $\omega_{SPP}$ is the resonant angular frequency (eV), $\Gamma_{tot}$ is the SPP total decay rate (eV) and is equal to the sum of absorption ($\Gamma_{abs}$) and radiative decay ($\Gamma_{rad}$) rates, $\delta$ is the in-coupling phase-shift, and $s_+$ is the amplitude of the incident wave power. A factor of $\cos\alpha_{min}$ is added to $s_+$ indicating only part of the input energy is coupled to SPPs. Since $a$ is harmonic with time, we solve Eq. (3) for $a = \dfrac{\sqrt{\Gamma_{rad}}\, e^{i\delta} \cos\alpha_{min}}{i(\omega - \omega_{SPP}) + \Gamma_{tot}/2} s_+$. If only the specular reflection is present so that the single port model is applicable and the SPP radiation is a linearly polarized but rotated from the incident plane by $\phi_{SPP} = \alpha_{min}$, the reflection coefficients of the parallel ($r_{para}$) and orthogonal ($r_{orth}$) components can then be expressed as [44,46]:

$$\begin{bmatrix} r_{para} \\ r_{orth} \end{bmatrix} = \begin{bmatrix} r_o + \dfrac{\Gamma_{rad} \cos^2 \alpha_{min} e^{i\varsigma}}{i(\omega - \omega_{SPP}) + \Gamma_{tot}/2} \\ \dfrac{\Gamma_{rad} \sin \alpha_{min} \cos \alpha_{min} e^{i\varsigma}}{i(\omega - \omega_{SPP}) + \Gamma_{tot}/2} \end{bmatrix}, \quad (4)$$

where $r_o$ is the non-resonant reflection background and $\varsigma$ is the total coupling phase-shift of SPPs and is close to zero for single port [46,49]. Here, the parallel and orthogonal components are defined as the analyzer is placed at $\gamma = 0°$ and $90°$. From Eq. (4), one sees the para- and orth-reflectivities are controlled by $\alpha_{min}$, which depends on the sample orientation, wavelength, and the mode order, and the interplay between the absorption and radiative decay rates of SPPs. For verification, we calculate the (-1,0) para- and orth- reflectivity spectra of the array and plot them in Fig. 7(a) for $\theta = 10°$ and several $\varphi$ under p-incidence. The parallel and orthogonal profiles appear as dips and peaks, respectively. The profiles are then fitted by Eq. (4) to determine $\Gamma_{rad}$, $\Gamma_{tot}$, and $\alpha_{min}$. The best fits are shown as the dash lines in Fig. 7(a) for comparison and the fitted results are plotted in Fig. 7(b) & (c) with $\lambda_{SPP}$. $r_o$ and $\varsigma$ are determined to be around -0.989 and -0.035, respectively, for all cases in Fig. 7(d).

To double check $\Gamma_{rad}$, $\Gamma_{abs}$, and $\alpha_{min}$, we independently calculate $\Gamma_{rad}$ and $\Gamma_{abs}$ under the same excitation conditions by using the time-domain method in Fig. 7(b) as described previously [46]. We also directly determine $\alpha_{min}$ in Fig. 7(c) by calculating the reflectivity as a function of $\alpha$ for each $\lambda_{SPP}$. Two methods show less than 4% discrepancy between CMT and direct calculation. Once the CMT model is ready, we attempt to reproduce the numerical results. The $\psi$ and $\chi$ spectra are calculated by using the deduced parameters and displayed in Fig. 7(e) & (f) together with the FDTD results. The consistency between the analytical and simulation results again echoes the CMT model.

## VI. Discussion

We are now in the position of interpreting the behaviors of $\psi$ and $\chi$ by using the CMT expressions. From Eq. (4), since $\zeta$ is close to zero (see Fig. 7(d)), the para- and orth-reflections are always in phase at $\lambda_{SPP}$, producing a linear polarization. However, when the wavelength is slightly off the resonance, the radiation of SPPs acquires an additional phase shift due to the imaginary term $i(\omega - \omega_{SPP})$ at the denominator. The nonresonant and the parallel component of the SPP radiations thus are no longer $\pi$ out of phase with each other. The interference between them then yields different elliptical polarization states depending on $i(\omega - \omega_{SPP})$, $\Gamma_{rad}$, $\Gamma_{abs}$, and $\alpha_{min}$.

From Fig. 7(e) & (f), we notice the $\psi$ and $\chi$ profiles at $\varphi = 18°$ (i.e. $\lambda_{SPP} = 952$ nm) deserve further attention. When scanning across the resonance, the polarization changes from almost right circularly polarized (i.e $\chi \sim -45°$) to orthogonal linearly polarized at $\lambda_{SPP}$ ($\chi = 0°$ and $\psi = \pm 90°$) and then to left circularly polarized ($\chi \sim 45°$) at longer wavelength. From Fig. 7(b) - (e), we find the fitted $r_o = -0.989$, $\alpha_{min} = 161.7°$, $\Gamma_{tot} = 5.64$ meV, $\Gamma_{rad} = 3.28$ meV, and $\zeta = 0.033$ give $r_{para} = 0.05$ and $r_{orth} = 0.136$ at 952 nm. The $r_{orth}/r_{para}$ ratio reaches 2.72, resulting in the orth/para reflectivity ratio = 7.4. In fact, the condition for achieving complete orthogonal polarization conversion can be understood by making $r_{para} = 0$ in Eq. (4), physically implying the nonresonant reflection is destructively interfered with the para-component of the SPP radiation. By assuming $r_o \sim -1$ and $\zeta \sim 0$, the condition $2\cos^2 \alpha_{min} - 1 = \Gamma_{abs}/\Gamma_{rad}$ would yield $r_{para} = 0$. In other words, for a given $\lambda_{SPP}$ so that $\Gamma_{abs}/\Gamma_{rad}$ is a constant, we may orient the sample to have $\alpha_{min}$ to facilitate complete para-to-orth polarization conversion. However, when $\Gamma_{abs}/\Gamma_{rad} = 1$, which signifies critical coupling, both $r_{para}$ and $r_{orth} = 0$, leading to total absorption [55].

A low $r_{orth}$ in this $\varphi = 18°$ case indicates much of the incidence energy is being lost to the absorption of SPPs. Useful polarization conversion requires not only $r_{para} = 0$ but at the same time $r_{orth} \sim 1$. To enhance $r_{orth}$, from Eq. (4), the array must be designed to have $\Gamma_{rad} \sin\alpha_{min} \cos\alpha_{min}$ being almost equal to close to $\Gamma_{tot}/2$ if $\zeta \sim 0$. Therefore, to fulfill two conditions simultaneously, $\Gamma_{rad}$ must be much larger than $\Gamma_{abs}$, making $\alpha_{min} = 45°$ or $135°$. It has been reported that under some circumstances where the hole size is smaller than the period, $\Gamma_{rad}$ follows the Rayleigh scattering of single isolated holes with $\left(R\sqrt{H}/\lambda\right)^4$ while $\Gamma_{abs}$ can be considered as plain metal Ohmic absorption, which is $\sim \omega\varepsilon_m''/(\varepsilon_m')^2 \left(\varepsilon_m'/\varepsilon_m' + 1\right)^{\frac{3}{2}}$, where $\varepsilon_m'$ and $\varepsilon_m''$ are the real and imaginary parts of the metal dielectric constant, $\Gamma_{abs}/\Gamma_{rad}$ could be much reduced by properly designing the geometry and the material of the system [56]. To illustrate that, we perform FDTD simulation on a Ag array as it has smaller $\Gamma_{abs}$ than that of Au at the optical wavelength. Our approach is as follows. We choose an array with P = 1600 nm, R = 640 nm, and H = 300 nm such that the hole diameter is as close to the period as possible for maximizing $\Gamma_{rad}$ while at the same time the $\Gamma_{abs}$ of the (-1,0) mode at near infrared is minimal. To roughly locate the sample orientation for $\alpha_{min} = 135°$, we calculate the dispersion relation by the phase matching equation and the plot of $\alpha_{min}$ with $\lambda_{SPP}$ in Fig. 8(a) & (b) at $\theta = 10°$. As indicated by the dash lines, $\varphi$ is close to 37.7° for $\alpha_{min} \sim 135°$. Fig. 8(c) then shows the FDTD calculated para- and orth-reflectivity spectra calculated at several $\varphi$ from 29° to 37° under p-excitation. Actually, at $\varphi = 33°$, para- and orth-reflectivities are found to be 0.053 and 0.963, respectively, at $\lambda_{SPP} = 1.842$ μm, leading to orth/para reflectivity ratio = 333. By fitting the spectra using Eq. (4), we find $\Gamma_{rad}$ and $\Gamma_{abs}$ = 11.44 and 0.44 meV and $\alpha_{min} = 136.38°$.

Finally, for the circular polarization, both r$_{para}$ and r$_{orth}$ should have comparable magnitude but retard in a relative phase of 90°. As aforementioned, at $\omega \neq \omega_{SPP}$, the reflection coefficients can be rewritten as $\begin{bmatrix} r_{para} \\ r_{orth} \end{bmatrix} = \begin{bmatrix} r_o + Ae^{i\kappa}\cos\alpha_{min} \\ Ae^{i\kappa}\sin\alpha_{min} \end{bmatrix}$, where A and κ are constants depending on $\alpha_{min}$, $\Gamma_{tot}$, $\Gamma_{rad}$, $\omega - \omega_{SPP}$, and ζ. Therefore, circular polarization requires $\frac{r_o + Ae^{i\kappa}\cos\alpha_{min}}{Ae^{i\kappa}\sin\alpha_{min}} = \pm i$. For the φ = 18° case, r$_{para}$ and r$_{orth}$ are found to be close to $-0.989 + \frac{0.0029}{i(\omega-\omega_{SPP})+0.00282}$ and $\frac{-0.000978}{i(\omega-\omega_{SPP})+0.00282}$ by taking ζ ~ 0. Therefore, their division is close to $\pm i$ when $\omega - \omega_{SPP} = \mp 0.00099 - 0.000169i$, which agrees with the results in Fig. 7(e) & (f) where χ ~ ±45° are found at $\omega - \omega_{SPP} = \mp 0.00097$.

## VII. CONCLUSION

In summary, we have studied polarization conversion from 2D Au nanohole arrays by angle- and polarization resolved reflectivity spectroscopy. Although the arrays do not possess any intrinsic chirality, both the rotation angle and ellipticity measurements have indicated Bloch-like propagating SPPs play a significant role in facilitating extrinsic chirality. The experimental, numerical, and analytical results reveal the interference between the nonresonant background and the SPP radiation manifests various polarization states ranging from linear to elliptical polarization across the SPP resonance. While the nonresonant background preserves the incident polarization, the properties of the SPP radiation are strongly dependent on the vectorial near field pattern of SPPs and the interplay between their absorption and radiative decay rates. As a result, by controlling the sample orientation and geometry to tailor the field pattern and decay rates, it is possible to achieve almost complete parallel to orthogonal linear and parallel to circular polarization conversions.

## VIII. ACKNOWLEDGEMENT

This research was supported by the Chinese University of Hong Kong through the RGC Competitive Earmarked Research Grants (402908, 403310, and 14304314), CRF CUHK1/CRF/12G, and AOE AoE/P-02/12.

**Figure Captions:**

1. (a) The schematic of the angle- and polarization-resolved reflectivity spectroscopy. The incident and azimuthal angles are defined as $\theta$ and $\varphi$, respectively. Insets: the plane-view SEM image of the Au array used for measurement and the cross-section image of the unit cell used for the FDTD simulation. The (b) experimental and (c) FDTD simulated contour p-polarized specular reflectivity mappings. The dash lines indicate the excitation of (-1,0) and (0,-1) SPP modes calculated by the phase-matching equation. At $\varphi = 45°$ where two SPP modes cross with each other, a plasmonic band gap occurs together with two hybridized bright and dark modes located at shorter and longer wavelengths.

2. (a) The schematic for measuring the four Stokes parameters of the specular reflection. $E_p$ and $E_s$ are defined as the p- and s-polarizations. The experimental (b) rotation angle $\psi$ and (c) ellipticity $\chi$ contour mappings taken at $\theta = 10°$ under p-excitation. Noticeable $\psi$ and $\chi$ are seen at the (-1,0) and (0,-1) SPP excitations. (d) The $\psi$ extracted along the (-1,0) and (0,-1) SPP modes. The dash line indicates $\varphi = 45°$ where the anomalous "oscillations" superimposed on the broad background are seen. The plots of the largest positive and negative $\chi$ as well as the $\chi$ exactly at $\lambda_{SPP}$ for the (e) (-1,0) and (f) (0,-1) SPP modes. For the positive and negative $\chi$, similar anomalies are observed at $\varphi = 45°$ given by the dash lines. The $\chi$ at $\lambda_{SPP}$ is almost equal to zero and is independent of $\varphi$.

3. The corresponding FDTD simulated (a) rotation angle $\psi$ and (b) ellipticity $\chi$ contour mappings. (c) The $\psi$ extracted along the (-1,0) and (0,-1) SPP modes. (d) The largest positive and negative $\chi$ as well as the $\chi$ exactly at $\lambda_{SPP}$ for the (0,-1) SPP modes. Similar anomalies are seen at the gap region.

4. (a) The plot of the normalized reflectivity a function of incident polarization angle $\alpha$ taken at $\theta = 10°$ and $\varphi = 10°$, corresponding to $\lambda_{SPP} = 950$ nm. The best excitation condition $\alpha_{min}$ is determined by fitting the data with a sinusoidal function as given by the solid line. $\alpha_{min} = 168°$ as indicated by the arrow. (b) More normalized reflectivity curves together with the best fits taken at $\theta = 10°$ and $\varphi = 20°, 40°, 60°$, and $80°$, corresponding to $\lambda_{SPP} = 944, 916, 876$, and $830$ nm. (c) The plot of $\alpha_{min}$ as a function of $\lambda_{SPP}$ for $\theta = 10°$. It is noted that $\alpha_{min}$ diverges to $180°$ and $90°$ at $\lambda_{SPP} = 910$ nm where $\varphi = 45°$ for the bright and dark modes. The solid line is the analytical model based on $\hat{e} \cdot (\hat{k}_{SPP} \times \hat{z}) = 0$, where the unit vectors of $\hat{e}$, $\hat{k}_{SPP}$, and $\hat{z}$ are defined in the inset. The $\hat{e}_p$ and $\hat{e}_s$ are the p- and s-polarization vectors. (d) The plot of $\alpha_{min}$ as a function of $\lambda_{SPP}$ for $\theta = 5°, 10°$, and $15°$ together with the analytical model. Data around the gap region is excluded.

5. (a) The FDTD simulated $\alpha_{min}$ as a function of $\lambda_{SPP}$ (square symbol) calculated at $\theta = 10°$. The $\alpha_{min}$ deduced from the analytical model by using the Poynting vector under the same excitation condition (circle symbol). Inset: the cross-section image of the FDTD unit cell. The Poynting vector mappings taken at $\theta = 10°$ and two $\varphi =$ (b) $30°$ and (c) $45°$ for $\lambda_{SPP} = 936$ and $907.5$ nm, which correspond to a nondegenerate and hybridized bright SPP modes.

6. (a) The normalized orthogonal reflectivity measured as a function of $\alpha$ at $\theta = 10°$ and $\varphi = 10°$ ($\lambda_{SPP} = 950$ nm). The solid line is the best fit for determining $\phi_{SPP}$. (b) More orthogonal reflectivity curves taken at $\theta = 10°$ and $\varphi = 20°, 40°, 60°$, and $80°$. The best fits are given by the solid lines. (c) The schematic for the developing the analytical mode for the $\phi_{SPP}$ determination. The polarization of the SPP radiation is defined by $\phi_{SPP}$ with respect to the incident plane. $\alpha + \gamma = 90°$ for the orthogonal polarizer-analyzer pair. (d) Comparison between $\alpha_{min}$ taken from Fig. 4(c) and $\phi_{SPP}$, showing $\alpha_{min} = \phi_{SPP}$.

7. (a) The FDTD simulated (-1,0) para- and orth-reflectivity spectra calculated for $\theta = 10°$ and $\varphi = 0°, 6°, 12°, 18°,$ and $24°$ under p-excitation, corresponding to $\lambda_{SPP} = 962, 961, 958, 952,$ and $945$ nm. The parallel spectra appear as dips whereas the orthogonal spectra are peaks. The solid lines are the best fits using the temporal CMT model. (b) The deduced $\Gamma_{rad}$ and $\Gamma_{abs}$ by using the CMT and the time domain methods for different $\lambda_{SPP}$. (c) The CMT deduced and the FDTD calculated $\alpha_{min}$ for different $\lambda_{SPP}$. (d) The deduced $r_o$ and $\zeta$ for different $\lambda_{SPP}$, showing they are almost constant at -1 and 0. Comparison between the CMT deduced (solid lines) and the FDTD simulated (symbols) (e) rotation angle $\psi$ and (e) ellipticity $\chi$ for $\theta = 10°$ and $\varphi = 0°, 6°, 12°, 18°,$ and $24°$.

8. (a) The dispersion relation of a Ag array calculated by the phase matching equation at $\theta = 10°$. (b) The plot of $\alpha_{min}$ with $\lambda_{SPP}$ in the analytical model at $\theta = 10°$. The $\varphi$ is determined to be ~ 37.7° by the dash lines for $\alpha_{min}$ ~ 135°. The FDTD calculated (-1,0) (c) para- and (d) orth-reflectivity spectra for $\theta = 10°$ and $\varphi = 29°, 31°, 33°, 35°$ and $37°$ under p-incidence. The parallel and orthogonal reflectivity spectra show as dips and peaks.

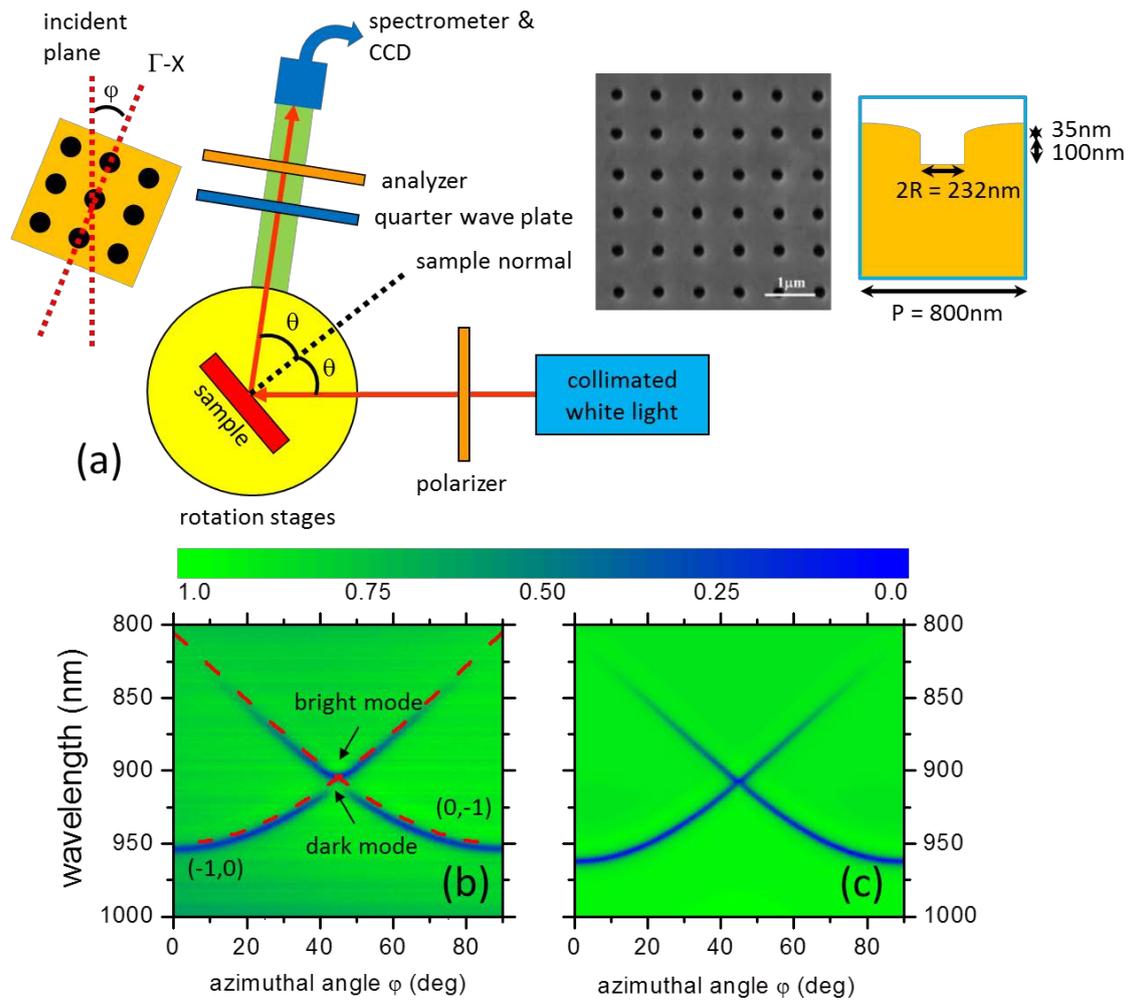

**Fig. 1**

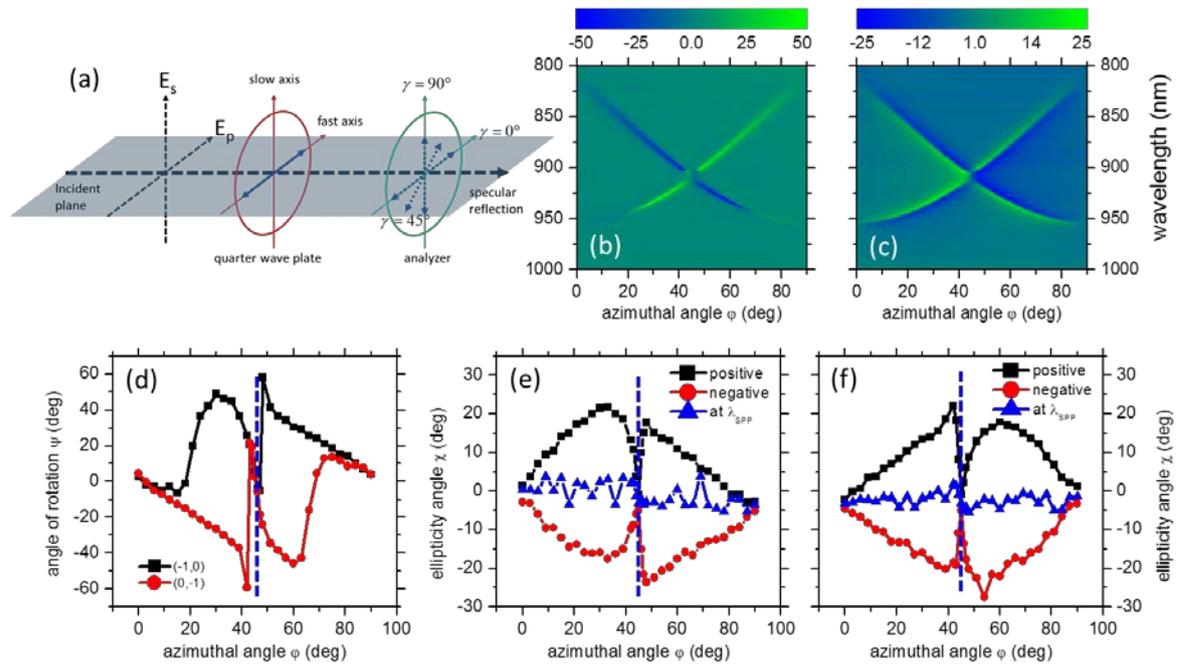

**Fig. 2**

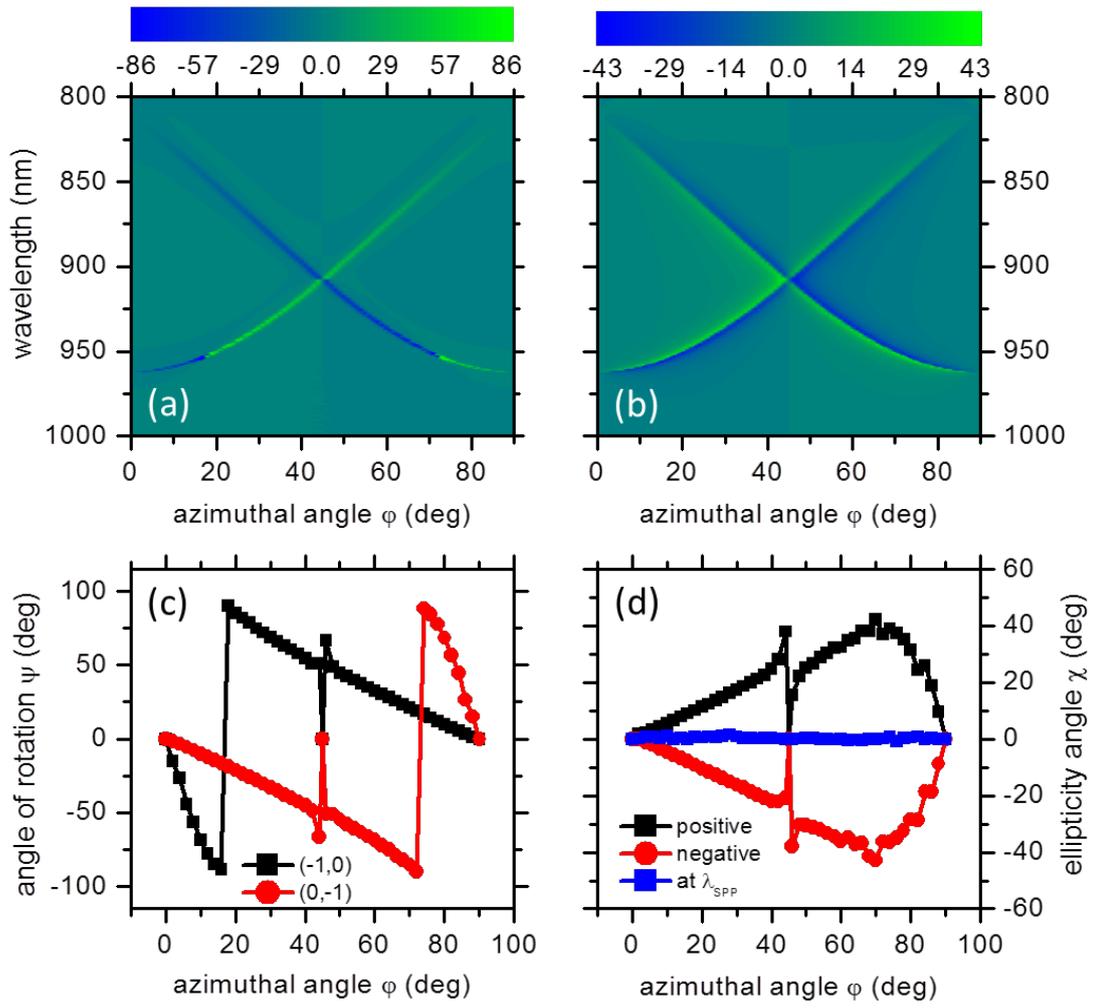

Fig. 3

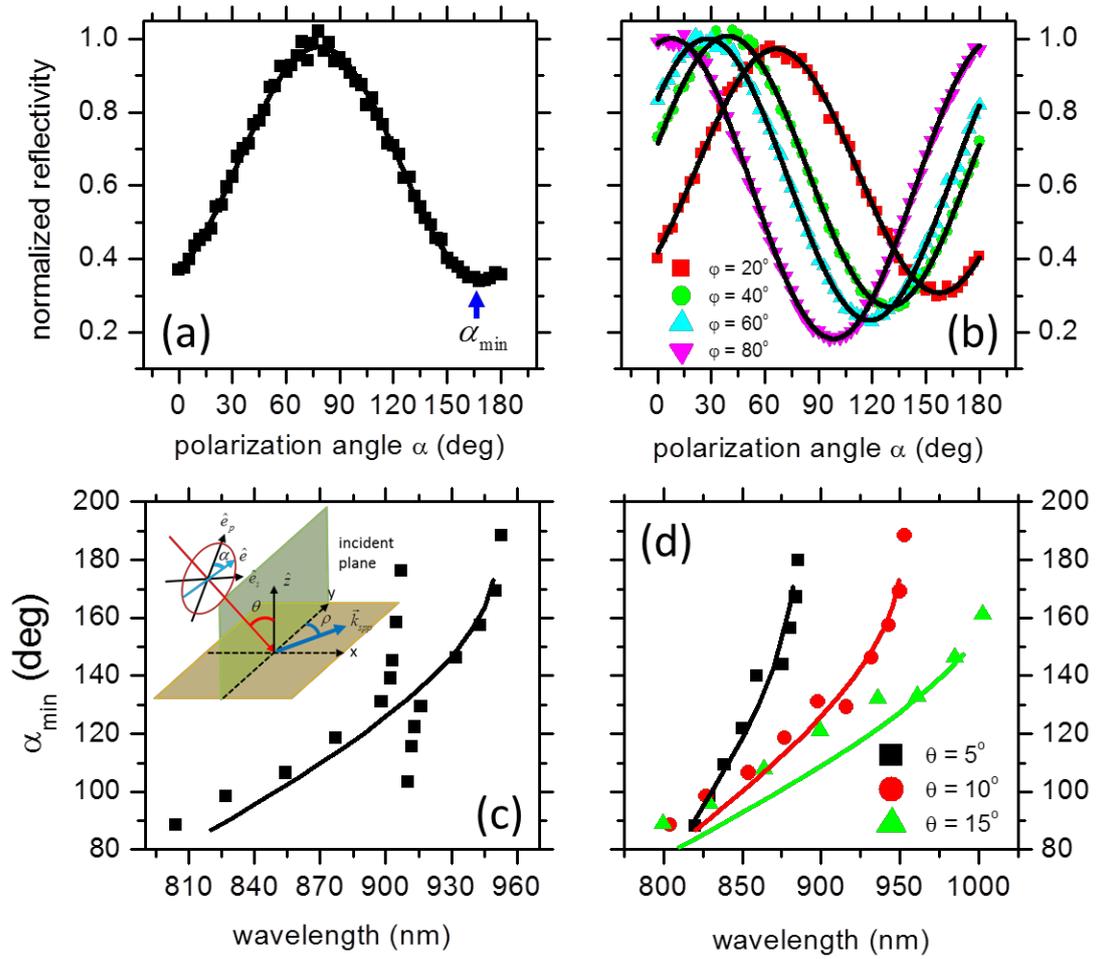

Fig. 4

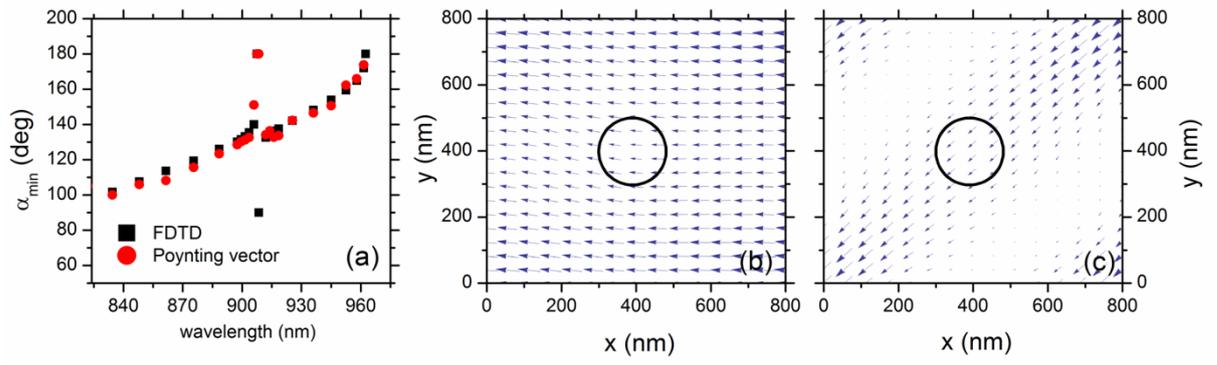

**Fig. 5**

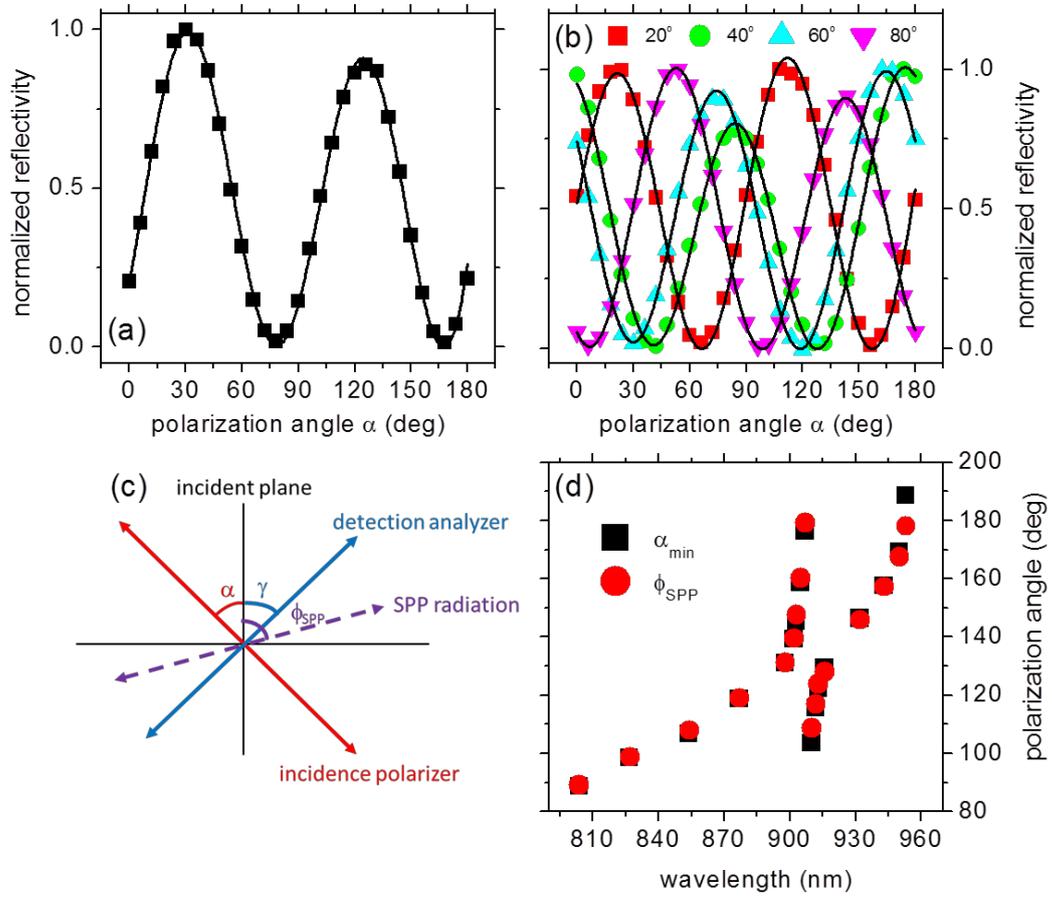

Fig. 6

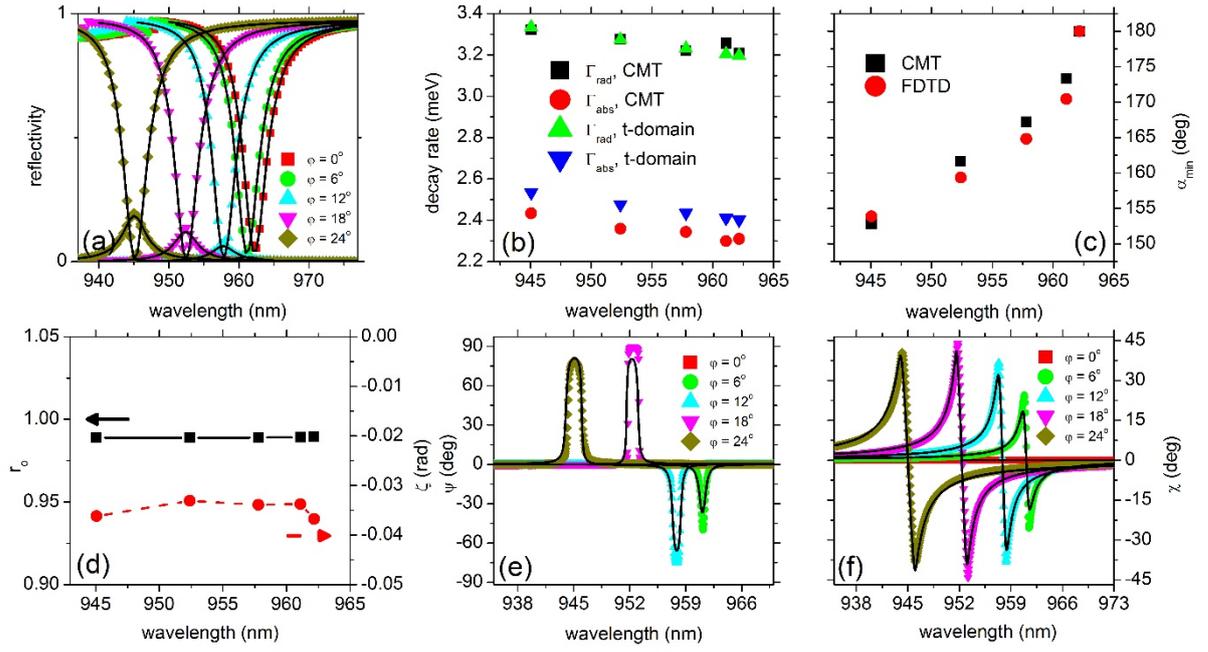

**Fig. 7**

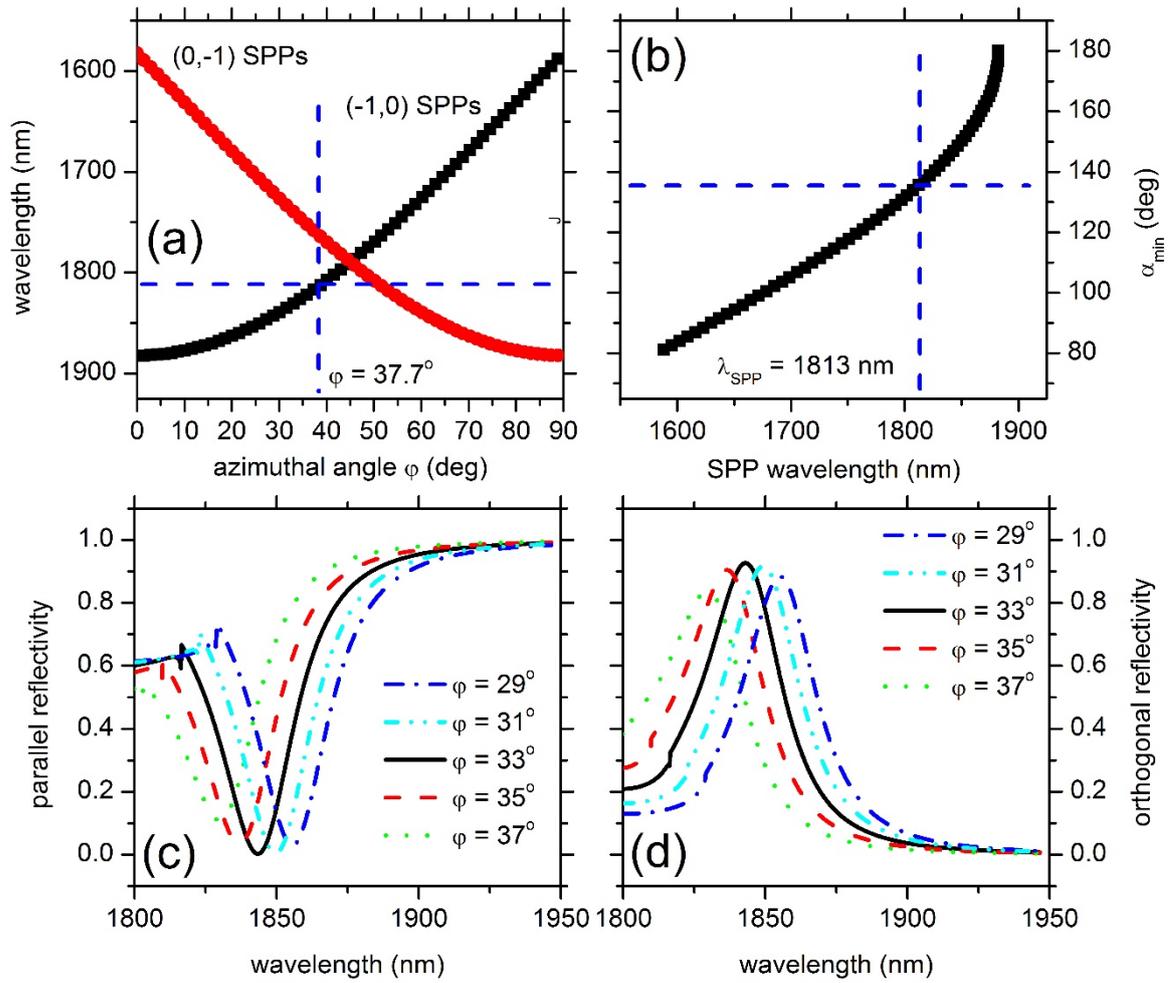

Fig. 8